\documentclass[journal]{IEEEtran}
\ifCLASSINFOpdf
\else
\fi
\usepackage{amsmath}
\usepackage{amsfonts}
\usepackage{graphicx}
\usepackage{graphicx}
\usepackage{xcolor}
\usepackage{cite}
\usepackage{epstopdf}
\usepackage{subcaption}
\usepackage[left=0.55in,right=0.55in,top=0.6in,bottom=0.6in, textheight=18in]{geometry}

\makeatletter
\def\BState{\State\hskip-\ALG@thistlm}
\makeatother

\begin{document}
\raggedbottom

%
% paper title
% Titles are generally capitalized except for words such as a, an, and, as,
% at, but, by, for, in, nor, of, on, or, the, to and up, which are usually
% not capitalized unless they are the first or last word of the title.
% Linebreaks \\ can be used within to get better formatting as desired.
% Do not put math or special symbols in the title.
\title{Control over Skies: Survivability, Coverage, and Mobility Laws for Hierarchical Aerial Base Stations}
% with support for Augmented Reality
%
% author names and IEEE memberships
% note positions of commas and nonbreaking spaces ( ~ ) LaTeX will not break
% a structure at a ~ so this keeps an author's name from being broken across
% of the two lines.
% use \thanks{} to gain access to the first footnote area
% a separate \thanks must be used for each paragraph as LaTeX2e's \thanks
% was not built to handle multiple paragraphs
%

\author{Vishal Sharma, Navuday Sharma, Mubashir Husain Rehmani, Haris Pervaiz
%\thanks{V. Sharma is with the School of Electronics, Electrical Engineering and Computer Science (EEECS), Queen's University Belfast (QUB), Northern Ireland, United Kingdom.}% <-this % stops a space
%\thanks{N. Sharma is with Ericsson, Estonia.}
%\thanks{M.H. Rehmani is with the Department of Computer Science, Cork Institute of Technology (CIT), Ireland.}
%\thanks{H. Pervaiz is with the School of Computing and Communications (SCC), Lancaster University, UK.}
}

% note the % following the last \IEEEmembership and also \thanks -
% these prevent an unwanted space from occurring between the last author name
% and the end of the author line. i.e., if you had this:
%
% \author{....last name \thanks{...} \thanks{...} }
%                     ^------------^------------^----Do not want these spaces!
%
% space would be appended to the last name and could cause every name on that
% line to be shifted left slightly. This is one of those "LaTeX things". For
% instance, "\textbf{A} \textbf{B}" will typeset as "AB" not "AB". To get
% "AB" then you have to do: "\textbf{A}\textbf{B}"
% \thanks is no different in this regard, so shield the last } of each \thanks
% that ends a line with a % and do not let a space in before the next \thanks.
% Spaces after \IEEEmembership other than the last one are OK (and needed) as
% you are supposed to have spaces between the names. For what it is worth,
% this is a minor point as most people would not even notice if the said evil
% space somehow managed to creep in.

% The paper headers
\markboth{}%
{}
% The only time the second header will appear is for the odd-numbered pages
% after the title page when using the two side option.
%
% *** Note that you probably will NOT want to include the author's ***
% *** name in the headers of peer review papers.                   ***
% You can use \ifCLASSOPTIONpeerreview for conditional compilation here if
% you desire.

% If you want to put a publisher's ID mark on the page you can do it like
%:
%\IEEEpubid{0000--0000/00\$00.00~\copyright~2015 IEEE}
% Remember, if you use this you must call \IEEEpubidadjcol in the second
% column for its text to clear the IEEEpubid mark.

% use for special paper notices
%\IEEEspecialpapernotice{(Invited Paper)}

% make the title area
\maketitle

% As a general rule, do not put math, special symbols or citations
% in the abstract or keywords.
\begin{abstract}
Aerial Base Stations (ABSs) have gained significant importance in the next generation of wireless networks for accommodating mobile ground users and flash crowds with high convenience and quality. However, to achieve an efficient ABS network, many factors pertaining to ABS flight, governing laws and information transmissions must be studied. In this article, multi-drone communications are studied in three major aspects, survivability, coverage, and mobility laws, which optimize the multi-tier ABS network to avoid issues related to inter-cell interference, deficient energy, frequent handovers, and lifetime. The article includes simulation results of hierarchical ABS allocations for handling a set of users over a defined geographical area. Several open issues and challenges are presented to provide deep insights into the ABS network management and its utility framework.
\end{abstract}

% Note that keywords are not normally used for peer review papers.

% For peer review papers, you can put extra information on the cover
% page as needed:
% \ifCLASSOPTIONpeerreview
% \begin{center} \bfseries EDICS Category: 3-BBND \end{center}
% \fi
%
% For peer review papers, this IEEEtran command inserts a page break and
% creates the second title. It will be ignored for other modes.
\IEEEpeerreviewmaketitle

\section{Introduction}
Due to the recent advancement of vehicular technology in 5G, Unmanned Aerial Vehicles (UAVs), commonly known as drones have gained a significant consideration to be used as \textcolor{black}{Aerial Base Station (ABS)} for facilitating cellular connectivity to ground mobile users~\cite{kaleem2018amateur}\cite{8255735}. To support a high density of users under flash crowd traffic, in the events of concerts, mass gatherings, cultural festivals, and sports, on-demand ABS deployments ensure offloading of traffic in Terrestrial Cellular Network (TCN) \cite{7932923}. Multi-tier aerial architecture complements TCN to serve the users under high shadowing and interference effects and is well-studied in \cite{8316776} that lists the challenges associated with the multi-tier aerial vehicles. \textcolor{black}{However, it does not provide any collective solution for the core requirements of mobility and coverage of ABS, which are the essential parts of the work presented in this article. Furthermore, the survivability that depends on the prediction and estimation of the lifetime of each UAV needs further exploration.}

\textcolor{black}{ABS network offers certain benefits over TCN, such as dynamic and adaptive cell coverage, being deployed as a flying relay with drone cells integrated with macro and microcells, where drone cell coverage can be changed by varying the power and altitude depending on the data traffic. However, severe interference from macro, micro, and other drone cells must be minimized using interference mitigation techniques and drone trajectory planning to avoid cell overlap. Moreover, the Total Cost of Ownership (TCO) of mobile operators is reduced by integrating the ABS network with TCN, since the energy requirements of ABSs are lower compared to terrestrial base stations, and site availability for cell planning is not required especially in the case of on-demand and public safety communications. However, the power consumption of ABS is impacted by the type of model, payload, capacity, and interference-management expected from their deployment, and there exists a power-supply issue when using these on-demand ABS, which needs far-more efficient solutions than the available technologies~\cite{jiang2020power}\cite{zeng2020federated}\cite{yan2020uav}.} %It must be followed that the ABS is not the only use case for drones in 5G. Drones play an important role in public safety networks used by military, police, fire, and emergency medical services in case of natural disasters, search and rescue operations, surveillance and reconnaissance\cite{7452263}\cite{7842423}.
\begin{figure*}[!ht]
 \centering
  \includegraphics[width=340px,trim = {1cm 2cm 0cm 1cm}, clip]{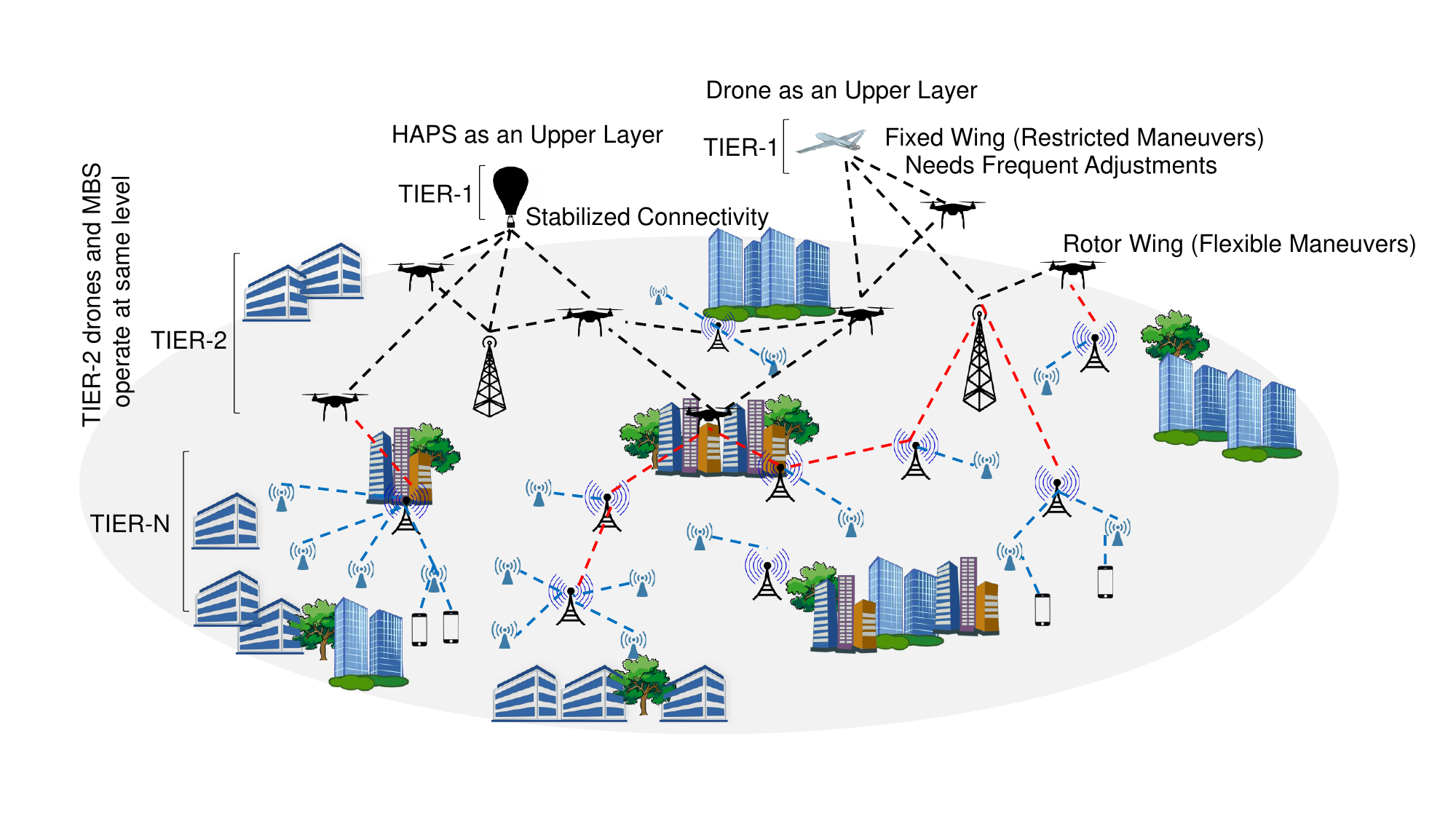}
  \caption{An exemplary illustration of hierarchical ABS setup for maximum coverage and agile reconfigurability~\cite{sharma2019neural}} %The multi-tier of drones facilitates easier network management and allows better service facilities to its users. Tier-2 drones and MBS are operated as the same network level with a similar set of instructions and capacities.}
  	\label{fig1}
\end{figure*}

\textcolor{black}{Artificial Intelligence} can be used as a powerful tool to reap the benefits of the ABS network by addressing several challenges such as efficient Xhaul and trajectory planning with cooperative and secured multi-drones data transmission with machine learning algorithms for predicting on-demand deployment of drone network. Another domain of research facilitated by the use of ABS is \textcolor{black}{mobile edge computing}-enabled \textcolor{black}{fog radio access network}, where functions such as signal processing and computing, resource management and allocation, distributed storing and caching abilities are performed at ABSs which act as a moving cloudlet. \textcolor{black}{However, to achieve aspects of the ABS network, well-defined coverage and mobility controls are of essential importance. There is very limited research for ABSs on these issues.}

In \cite{6881648}, the authors propose a distributed control algorithm for unmanned aerial and ground vehicles to perform desired tasks with minimum cost functions. The cost function for each agent or vehicle is different and results are validated with experimental tests. In \cite{6303888}, the authors address a similar problem with different cost function which is based on the health state of aerial vehicles. Many authors have addressed the problem of coverage and mobility control in wireless sensor networks based on different objective and cost functions. \textcolor{black}{The resulting questions include understanding the suitability of multi-tier ABSs architecture in the next generation of wireless networks. Alongside this, research on constraints, optimization issues and criteria for mobility, coverage, and survivability aspects of multi-tier ABSs is the leading concern identified from the existing works. To the best of the authors' knowledge, this article is unique in its approach for collectively focusing on the utility of multi-tier ABSs.}

%The rest of the article is organized as follows. Section II describes the hierarchical architecture of the system with the ABS network integrated with TCN. In Sec. III, problem description, and scope of multi-tier system architecture are defined. %%Aerial Base Station (ABS): Utilities and Extensions
\subsection{Multi-Tier ABSs and Hierarchy}
Multi-tier ABSs allow the use of drones as Macro Base Station (MBS) to facilitate the connectivity to the users with a similar capacity to that of a traditional MBS. There exists a plethora of approaches that fixates on a single layer of drones to enhance the Quality of Service/Quality of Experience (QoS/QoE) for end users while using drones as \textcolor{black}{access points}. Such solutions have been widely accepted because of their theoretical idealizations. The amalgamation of drones and traditional setup is not that convenient as assumed by the existing solutions. \textcolor{black}{To counterfeit such a challenge, it is suggested by different researchers and organizations to use drones as a network component and manage traffic by using drones like a normal network node. However, there are no concurrent studies that involve the evaluation of drones' behavior as well as its properties and maneuverability while deploying them in the network.}
%\footnote{Most of the existing studies assumes the flying capabilities and mobility management of drones}

In order to reduce the considerable impact on TCO, this article recommends using drones in multiple layers like TCN. As illustrated in Fig.~\ref{fig1}, N-layers can be formed by incorporating drones in TCN with Tier-2 drones acting similar to MBS, and High-Altitude Platform System (HAPS) and Tier-1 drones facilitate the movement and control over the underlying network. \textcolor{black}{This type of deployment offers a wide range of flexibility in allocating load and mobility management. The user-load from the access points is handled by Tier-2 drones where the Tier-1 drones offer backbone solution to this type of network}. It is to be noted that the type and the make of drones pose a considerable effect on the performance of the network as it is easier to regulate the network with rotor-wing drones, whereas fixed-wing drones require specialized algorithms for generating the waypoints.
\subsection{\textcolor{black}{Single-Tier Deployment vs. Multi-Tier Deployment}}
\textcolor{black}{Single-tier ABSs are easier to manage, control and operate irrespective of the scalability in terms of the number of drones functional at the same time over a specified area. In contrast to this, multi-tier ABSs help to define a new set of network architecture with a wide range of capacity, coverage, and operations, but with a complex formulation. Such layered architecture, if optimized successfully, offers many applications through elongated connectivity\cite{sharma2019neural}. Some of the applications of multi-tier ABSs include ultra-dense formation (high-density) for supporting urban networks with better coverage~\cite{8316776}\cite{sharma2019neural}, precision agriculture and city traffic monitoring~\cite{xu2020two}.}

\textcolor{black}{Multi-tier ABSs suffer from the critical issues of survivability, coverage enhancement, and mobility management. The problem with survivability is dependent on the resource depletion of drones, which leads to its failure and non-functioning after a period. The problem with the coverage is related to the positioning of drones and network planning, which causes issues related to fading, interference, and signal distortion. The problem with mobility is related to shifting of services between the drones and allocation of resources, failure of which leads to an isolated network with increased overheads. Thus, it is desired to design a multi-tier ABSs because of their capabilities but with a resolution of issues related to survivability, coverage, and mobility management. The contributions of this work include,
\begin{itemize}
    \item a discussion on several research concerns and constraints related to the laid requirements of multi-tier ABSs.
    \item exploring survivability, coverage, and mobility laws for multi-tier ABSs with a novel framework for optimizing the probability of connectivity and likelihood of mapping between the ABSs and the demand areas.
    \item a simulation case study is presented, which shows significant gain while supporting communication between the ABS, designated TCN-MBS, and a set of users in a defined geographical area.
\end{itemize}}
%\subsection{Scope of this article}%%%uncomment if necessary
%This paper considers the scenario of multi-tier ABSs while fixating a solution for optimized positioning and maneuvering of drones for maximizing the probability of connectivity and likelihood of mapping between the ABSs and the demand areas through a novel N-block recursive learning (NBRL) framework. In addition, this paper discusses the optimization issues and constraints related to the laid requirements of multi-tier ABSs. Moreover, a simulation case study is presented, which shows significant gains observed for the different sets of metrics while supporting communications between the ABS, designated TCN-MBS and a set of users in a defined geographical area.

\begin{figure*}[!ht]
  \centering
  \includegraphics[width=280px]{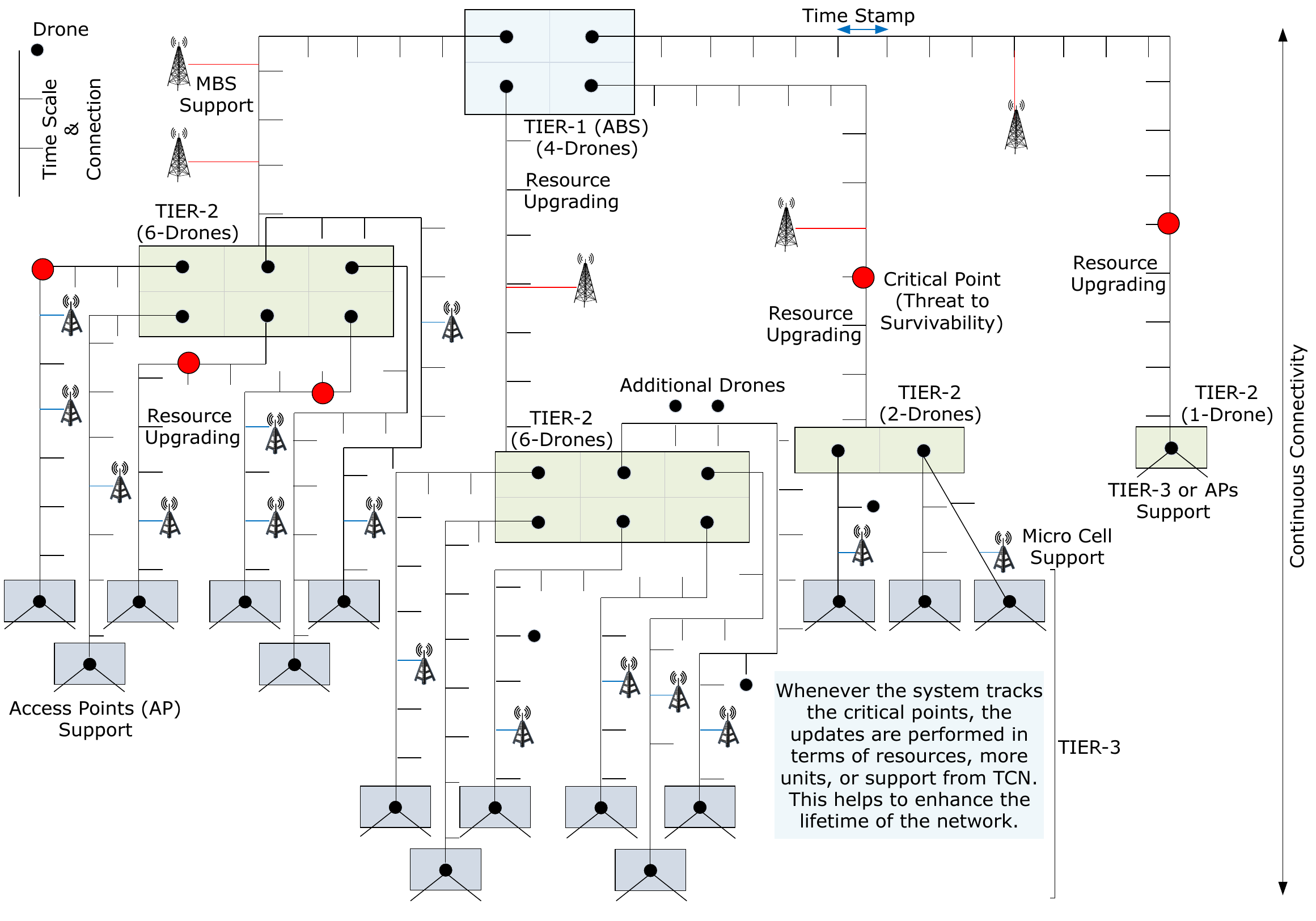}
  \caption{An illustration of the matrix-timing diagram for tracking the survivability of the system and deciding the requirements for additional resources, entities or support in terms of load balancing. }\label{fig2}
\end{figure*}

\section{\textcolor{black}{Research Concerns} of Multi-Tier ABSs}
\textcolor{black}{Several research concerns as listed below must be resolved to fulfil the functional requirements of multi-tier ABSs:}
\subsection{Dynamic Cell Coverage Area and Trajectory Planning}
One of the interesting features of the ABS network over TCN is its dynamically changing cellular coverage which depends mainly on ABS's altitude, power, and propagation environment. Reference \cite{6863654} studies the optimal altitude of ABS in Suburban, Urban, Dense Urban, and Urban High Rise environments for maximum cell coverage. Also, the probability of Line-of-Sight (LoS) is an important factor for an ABS which depends on the type of propagation environment and elevation angle between the ABS and the ground user. A closed-form expression for the probability of LoS is also provided in \cite{6863654}. \textcolor{black}{\textit{To obtain appropriate coverage control, the major design constraints that need to be optimized are ABS altitude and power.}} However, in a multi-tier architecture, ABS cell coverage would be affected by the inter-cell interference from other ABS cells, microcells, and macrocells. Therefore, appropriate trajectory planning of ABS is needed to avoid cell overlaps considering the drones' flying speed, direction, and acceleration.
\subsection{Energy Consumption}
ABSs are energy critical machines, therefore alternative sources of power such as mounted solar panels, stop and recharge techniques, radio frequency power sources are currently being researched and tested by deployment\cite{7888557}. The energy consumption by drones is divided into two functions- drone flight with onboard processing and information transmission by mounted Long Term Evolution (LTE) base station. \textcolor{black}{The power required for on-drone flight is higher than the transmission power. However, the power needed for fixed-wing aircraft is lower than the rotary ones of the same size and payload carrying capacity, but rotary wing drones can provide better coverage with lower hand-offs and Doppler effects due to their hovering capability at different altitudes whereas fixed-wing drones need to fly at velocity above a certain threshold to maintain stable fight, which causes higher Doppler and hand-offs.}
\subsection{ABS Placement}
ABS placement depends mainly on the user density requesting for services and its power. However, some other factors affecting the placement includes the probability of LoS with the ground mobile user, type of propagation environment, and implied collision avoidance techniques with other ABS or surrounding infrastructure. With a priori placement decision, ABSs discover the optimal trajectory and altitude towards the required set of ground users. Therefore, optimal ABS placement is needed to achieve coverage and mobility control.

%\subsection{Fast Handovers}
%Handover is another crucial optimization concern with ABSs. As described in the previous section, fixed-wing ABSs tend to have higher handovers than rotary wing ABSs. Since ABSs provide better coverage, it is easier to consider existing LTE handover standards for managing handoffs that arise from the high mobility of users. Software Defined Networking (SDN) can be an efficient solution to handle handovers, allowing for efficient control, management, and cooperation between the ABSs and the users. Also, there is a major requirement for handling vertical handovers if one considers the differences in the technologies used by the underlying infrastructure. For handover management between the infrastructure and the drone, successful techniques have been performed by Qualcomm in their practical tests using high and low power terrestrial BS by forming macrocell and microcells, respectively.
\subsection{Resource Allocation and Management}
Appropriate allocation and management of the radio resources ensure the survivability of the drone. The resources addressed herein are for both Physical and Network layers such as available spectrum, \textcolor{black}{physical resource blocks} of LTE-grid, channel state information, and capacity. Apart from resources required to accomplish the ABS transmission and reception, resources are also needed to achieve successful flight and maneuvers which depends on the energy efficiency and waypoint prediction for optimal trajectory. %Also, the coordination of the ABS network is needed to prevent the use of the same PHY and NET resources for interference mitigation.

\section{\textcolor{black}{Survivability, Coverage, and Mobility Laws}}
Despite the advantages of ABS in the next generation of wireless networks, there are certain limitations on their full-fledged use, which include, \textit{control over the movement of the ABS by identification of waypoints, identification of appropriate location to increase the view over geographical area, number of users to be shifted over ABS, and identification of the number of ABS to perform a particular task \cite{7859281}.} To resolve the above-described limitation, this section introduces the concepts of survivability, coverage, and mobility laws for efficient localization of multi-ABSs and non-failure-based network formation for supporting QoS to the end users. In addition, the N-Block Recursive Learning (NBRL) framework is proposed, which considers the derived policies on survivability, coverage, and mobility for fixating the final decisions on the operations of hierarchical ABSs.

\subsection{Survivability}
The survivability is defined through resource-based policing, which depends on the prediction and estimation of the lifetime of each drone based on its maneuverability as well as consumption of available resources. At first, the network is assumed to operate with a set $S$ of MBS whose coverage is to be improved by deploying a set $D$ of drones in $N$ tiers. \textcolor{black}{Now, the survivability of the system is obtained by modifying the Lusser's formula~\cite{kopp1996system}, according to which, the survivability of hierarchical ABSs can be obtained as $\mathcal{S}_{\mathcal{T}}^{(t)}=\prod\limits_{i=1}^{N} \left(\mathcal{S}_{\mathcal{L}} . \frac{\mathcal{D}_{A}}{|D|}\right)_{i}$, where $\mathcal{D}_{A}$ is the active drones in a given layer, and $\mathcal{S}_{\mathcal{L}}=\prod\limits_{j=1}^{|D_{i}|} \left(\mathcal{S}_{\mathcal{D}} . \frac{\mathcal{C}_{A}}{\mathcal{C}_{T}}\right)_{j}$, $\mathcal{T}$ is the total operational time, $\mathcal{C}_{T}$ is the total connections and $\mathcal{C}_{A}$ is active connections between the nodes expressed over a normalized function $f_{t}\left( \beta, \mathcal{E}, \mathcal{R}, \tau, \lambda \right)$. Here, $\beta$ and $\mathcal{E}$ denote the memory and energy associated with each drone, respectively, $\mathcal{R}$ is the radio range, $\tau$ is the transmission time, and $\lambda$ is the mean user distribution. $\mathcal{S}_{\mathcal{D}}(t)$ is calculated over same the same function as $\frac{-1}{t} \log\left(\frac{f_t}{f_{0}}\right)$, where $t\;\leq\;\mathcal{T}$.}

The details on the survivability can be obtained by following the illustration presented in Fig.~\ref{fig2}. \textcolor{black}{The figure shows the $n\times m$ ($n,\;m \;\geq\;1$) matrix-timing tree for each aerial node and their corresponding interacting node in the network. The matrix-timing diagram helps to understand the impact of a node on the functionality of the entire network as well as it can be used for enhancing the maximum lifetime of the network by controlling the survivability inputs at any instance. Tier-1 ABS are used to regulate to timing-scale that maintains a constant check on the resource consumption also enables resource upgrading whenever the survivability of the network is degraded. The instances of timing diagram use several critical points to identify any threats to survivability. These critical points are set based on the time to recover or upgrade the resources for continuity of the network. Whenever the system tracks a critical point in the setup, the system undergoes updates that cover resources from TCN to improve lifetime. The timing scale can be set in the initial configuration of the network and checkpoints can either be pre-decided along with the limits on the resources or can be altered based on the area-load in the network.}
%\begin{figure}[!ht]
  %  \centering
  %  \includegraphics[width=150px]{s1}
  %  \caption{An exemplary illustration of Voronoi-based area division as observed by an aerial node. The four types of areas considered in the zone under communication are marked by a variation in color.}\label{voronoi}
%\end{figure}

\subsection{Coverage Control}
For coverage laws, a Voronoi-based strategy, inspired by C\'{o}rtes et al.~\cite{cortes2004coverage}, is considered, which aims at the formation of control laws for ABS. The proposed approach utilizes the centroid, mass, and polar moment of inertia to allow efficient placement and coordination of multiple ABSs as expressed in~\cite{cortes2004coverage}, however w.r.t. the movement of multi-ABSs. The 3D placements, as well as controller selection, are performed to keep a check on the mobility of ABS through a layered module. Note that ABSs in Tier-1 are responsible for most calculations and sharing the details with the underlying ABSs (Tier-2, Tier-3, \dots, Tier-N). The individual evaluations are dominated only in the case of isolations.
%A polytope is a finite set of points which are always bounded by flat sides. A convex polytope is the one such as a line segment joining any two points in the geometry. The coverage model is developed for a single ABS scenario, which is then extended to the entire network.

For the geographical division of the area, the region under ABS is marked by Voronoi constellations denoted by a convex polytope. Coverage control is obtained by location and placement optimization of the ABS, which can be attained by managing the polytope divisions and allocating ABS according to their physical properties. The polytope operates over 2D coordinates of geographical areas ($x$, $y$), which are obtained by marking the 3D location of a drone to its corresponding 2D point on the ground. The polytope constellations for entire area are obtained as the union of sub-polytopes, which denote the number of divisions of the underlying area. The placement of ABS in the entire polytope is controlled by a location polytope, which is also a convex polytope. This mapping between the location and area is obtained by matching polytope-points. However, it is difficult to perfectly match each drone to the desired location. There may exist an error in the location of drone, which is marked by some error correction $\epsilon$ such that the actual location of a drone for $L_{1}$ can be assumed to be $L_{1}\pm \epsilon_{1}$. Considering this mapping, the number of users around the MBS region is marked as high demand area, medium demand area, low demand area, and no demand areas, and Voronoi set $V$ can be written as $V=\{v_{1}, v_{2}, \dots, v_{k}\}$, where $v_{k}$ denotes the sub-mapping between the drones and the $k$th demand area. %%The location marking from these sets are determined by using the strategy by C\'{o}rtes et al.~\cite{cortes2004coverage}, however, by re-planning their Gaussian function with the area under displacement generated by the movement of each UAV.
\begin{figure*}[!ht]
  \centering
  \includegraphics[width=280px]{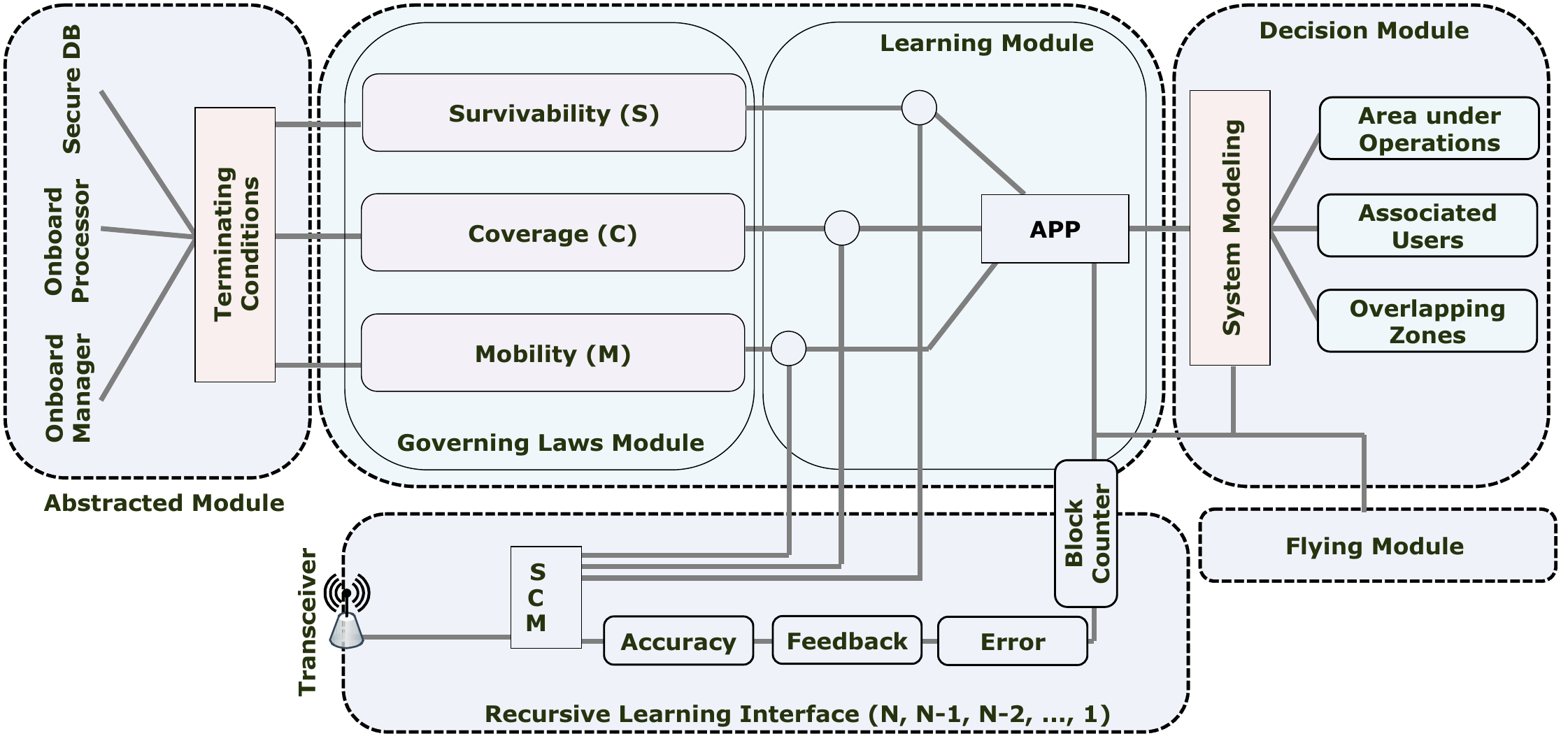}
  \caption{An illustration of the N-Block Recursive Learning Framework (NBRL) used for updating the policies and generating governing laws for survivability, coverage, and mobility.} %This model can be implemented on the Tier-1 ABS (or any central entity), which manages the calculations and passes on the information to connected ABS. The model can be implemented on each UAV also; however, such a deployment may consume excessive energy.}
  	\label{nbrl}
\end{figure*}
%%As performed earlier, the mobility laws are also defined for a single Tier-1 ABS and then extended to its corresponding ABS in the same tier.
\subsection{Mobility Laws}
The mobility laws are operated over control laws by considering the survivability of multiple drones as well as their operational hierarchy. For mobility laws, the movement of drones is managed through optimal placement based on Voronoi constellations. The mobility laws remain similar in all the tiers of drones irrespective of their functionality.%%% However, the primary aim of connectivity is maintained throughout the movement. This helps to avoid issues related to network partitioning as well as isolations.
%\footnote{A single Voronoi division can have multiple ABS depending on the area to be covered and the maneuverability of ABS.}
At first, the ABS are entitled to move according to their survivability factors and the request from the area, which is marked by the demand zones of Voronoi. The demand zones of Voronoi provide a non-overlapping placement of ABS along with the generation of estimated waypoints to control their movement. The calculation of area covered by each of the deployed UAV allows the removal of possible overlaps with an aim of maximum coverage with the minimum number of ABS. The Tier-1 ABS is coordinated by its serving TCN-MBS, whereas the next tier ABSs are coordinated by their previous tier ABSs, and in the case of NLoS or absence of coordinating node, the available ground infrastructure supports the coordination.

The mobility laws are derived considering two ABSs, which can be easily extended to $|\mathcal{D}|$ number of ABSs~\cite{sharma2019neural}. Let $D_{i}$ and $D_{j}$ be the two ABSs deployed with three different possibilities of no overlapping, completely overlapping with difference in altitude, and the partial overlapping. Since, the waypoints of ABSs are decided by their corresponding ABSs, TCN-MBS or any available infrastructure, complete overlap of two or more ABSs is not observed throughout this deployment. However, there can be a case of partial overlap which also affects the coverage area and causes interference, and increases the number of ABSs required to cover the entire MBS zone. Thus, overlapping of ABSs ($O_{A}$) irrespective of their tiers can be represented as a value between 0 and 1. To set mobility laws, strategy can be determined by using~\cite{sharma2019neural}, in which, the limits are defined for the total area covered by each of the two ABSs (with a sufficient difference in altitude) by using $x$ and $y$ coordinates. For this, the overlaps ($X_{o}$, $Y_{o}$) for the $x$ and the $y$ coordinates, respectively, can be calculated trivially, and based on the limits, $O_{A}$ can expressed as a Boolean, i.e. $1$ for $X_{o} > 0$ and $Y_{o} > 0$, which refers to overlapped movements, and $0$ otherwise, which refers to non-overlapped movements of ABSs in a single tier. %%%Similar formulations can be extended for expanding mobility laws to inter-tier ABSs.
%%%%%.is calculated as $X_{o} = \max (0, \min (D_{i}.x_{\max}, D_{j}.x_{\max})- \max(D_{i}.x_{\min}, D_{j}.x_{\min}))$, and $ Y_{o} = \max (0, \min (D_{i}.y_{\max}, D_{j}.y_{\max})- \max(D_{i}.y_{\min}, D_{j}.y_{\min}))$. Using these,
\subsection{\textcolor{black}{N-Block Recursive Learning (NBRL) Framework}}
This paper introduces the NBRL framework, which helps to provide update policies for survivability, coverage, and mobility laws, \textcolor{black}{as shown in Fig.~\ref{nbrl}. The framework helps to get periodic information from the N-Tiers in a recursive manner until the required criteria of operations are not satisfied. Here, the required criteria refer to the governing conditions for different laws associated with the successful operations of multi-tier ABSs. NBRL is a block-based framework extendable for any number of metrics. The abstracted module includes onboard devices and sensors, learning module is responsible for the governing laws on survivability, coverage, and mobility,  decision module performs the system modelling over the area under operations, associated users and overlapping zones based on Voronoi, which are monitored through the flying model and finally, recursive learning helps to improve the accuracy, ensure feedback and remove errors using block counters. However, the system requires parsing for evaluating the new set of policies from other nodes in the same or different tiers. Moreover, the NBRL framework accounts for the initial positioning of ABSs based on demand area to Voronoi mapping and checks for the covered area along with mobility management of ABS in the range specified for their tier. This framework considers the area covered by the entire fleet of ABSs in a single zone, and then, checks for coverage control, survivability options, and mobility laws.} In the case of maximum coverage and maximum survivability, the setup continues, whereas, in the case of non-mapping of the demand areas, the ABSs reshuffling and Voronoi re-mapping are performed to optimize the UAV placement and to maximize services without impacting the mobility laws. The framework is recursive and obtains its update in $N, N-1, N-2, \dots, 1$ pattern, which is iterated until the conditions selected for operations of ABSs are not close enough to the maximum likelihood for the associated laws.
%\footnote{The iterations also account for an individual in each layer, which refers to the available TCN or companion drones.}

\section{Performance Evaluation}
The proposed approach is evaluated numerically using a sample network setting in $Matlab^{TM}$. The analyses are carried in an area of 2500x2500 $m^{2}$ with each MBS having one active Tier-1 ABS with a communication range of 1000 m. The total number of tiers is set at 2, with Tier-2 ABSs serving like access points. The maximum number of Tier-2 ABSs used by the proposed approach is set to 20 over 1000 $m^{2}$ of area. The number of users varied between 1000 to 2000 per MBS region and each user made a service request using Poisson distribution with $\lambda$ varying between 5 and 10. The flying range of Tier-2 ABSs is set between 200 feet and 500 feet with the theoretic constraints of the Free Space Propagation model.

\begin{figure}[!ht]
\begin{subfigure}[b]{0.5\linewidth}
    \centering
   \includegraphics[width=140px]{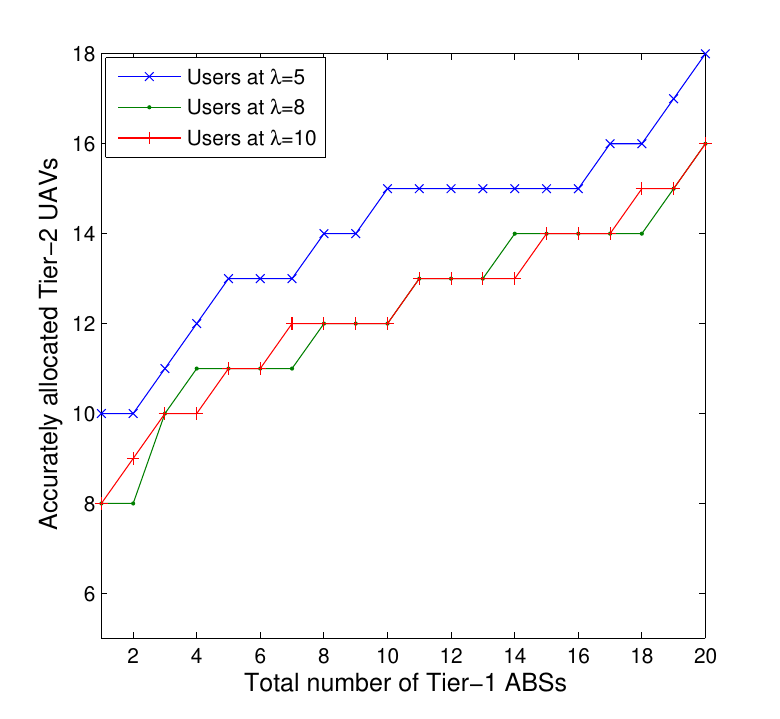}
    \caption{}\label{g1}
\end{subfigure}
%\hspace{0.1cm}
\begin{subfigure}[b]{0.5\linewidth}
    \centering
   \includegraphics[width=140px]{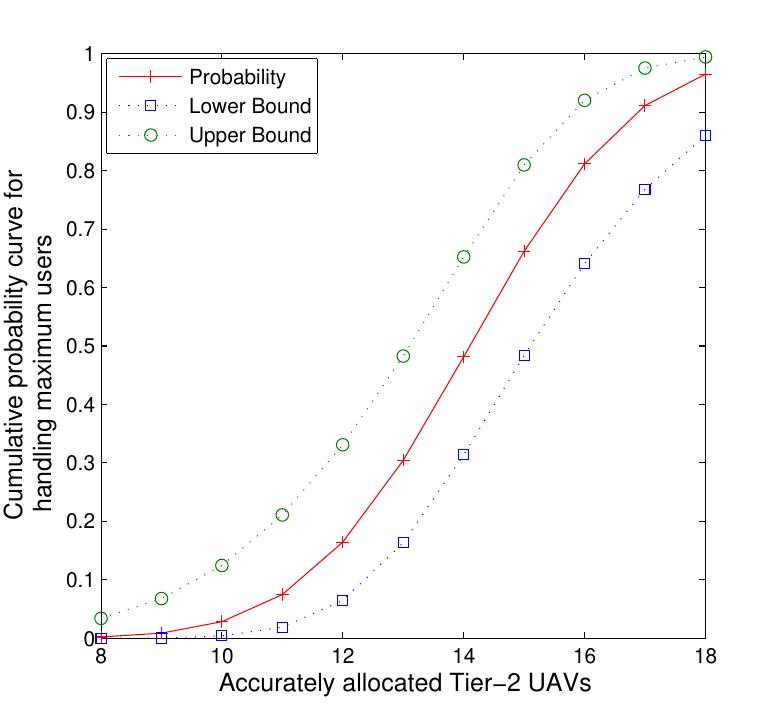}
    \caption{}\label{g2}
\end{subfigure}
\hspace{0.1cm}
\begin{subfigure}[b]{0.5\linewidth}
    \centering
  \includegraphics[width=140px]{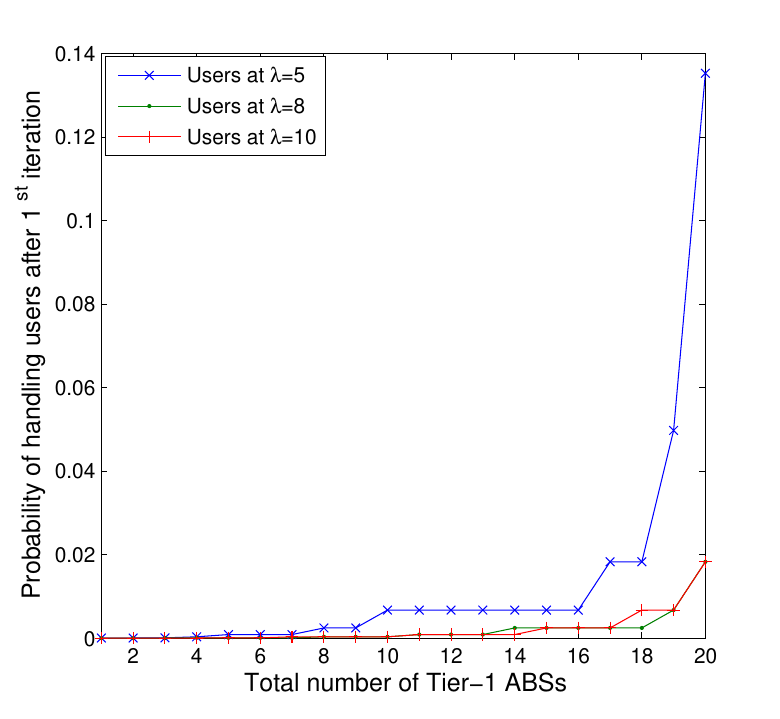}
    \caption{}\label{g3}
\end{subfigure}
%\hspace{0.1cm}
\begin{subfigure}[b]{0.5\linewidth}
    \centering
 \includegraphics[width=140px]{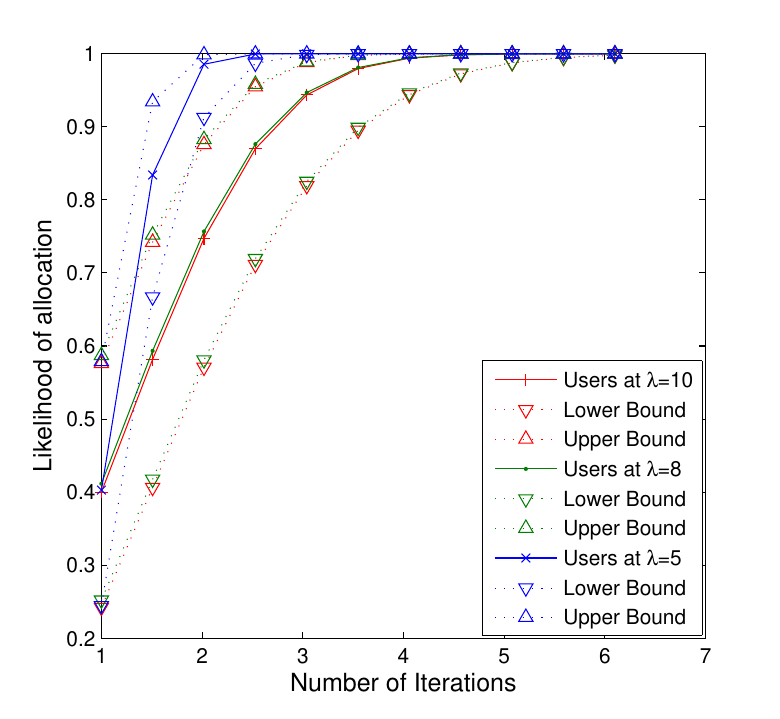}
    \caption{}\label{g4}
\end{subfigure}
%\vspace{0.1cm}
%\begin{subfigure}[b]{1\linewidth}
%    \centering
 %\includegraphics[width=240px]{g6}
 %   \caption{}\label{g5}
%\end{subfigure}
%
\caption{Simulation study (a) Accurately allocated Tier-2 ABSs vs. the total number of Tier-1 ABSs. (b) Cumulative probability for handling maximum users vs. The total number of accurately allocated Tier-2 ABSs. (c) Probability of handling users after $1^{st}$ iteration vs. The total number of Tier-1 ABSs. (d) Likelihood of ABSs allocation vs. Total iterations.}\label{fig_sim}
\end{figure}
\begin{figure}[!ht]
  \centering
  \includegraphics[width=170px]{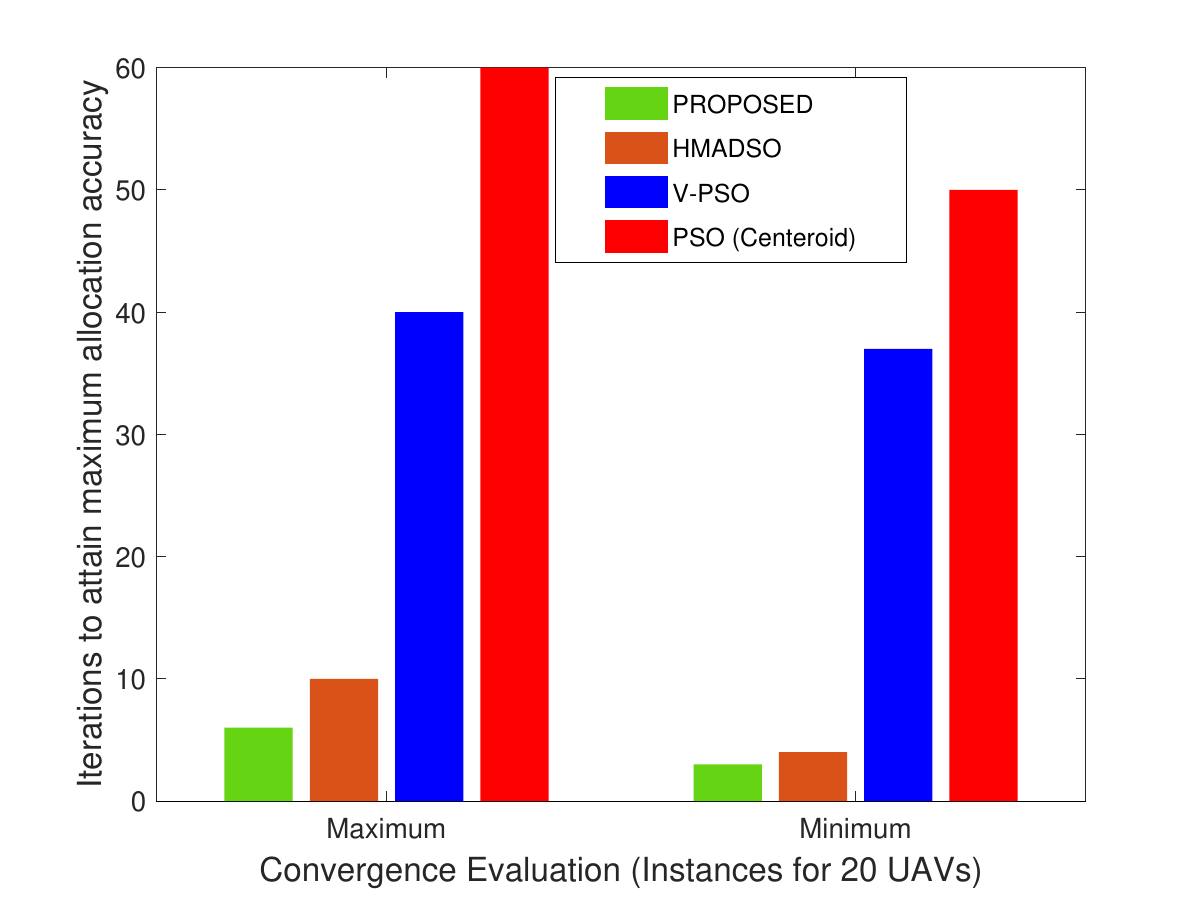}
  \caption{\textcolor{black}{Comparison between the proposed approach, HMADSO, PSO, and V-PSO for the number of iterations required to converge at an optimal solution.}}\label{fig_g6}
\end{figure}
%\footnote{The proposed approach is not tested for communication errors, rather survivability, coverage control, and mobility laws are tested.}
The proposed approach is tested for accuracy in allocating Tier-2 ABSs w.r.t. Tier-1 ABSs to their respective zones using location optimization of Voronoi, as shown in Fig.~\ref{g1}. The results show that the proposed approach can allocate Tier-2 ABSs with accuracy varying between 90\% and 40\% depending on the arrival rate of service requests and iterations as well as the available number of Tier-1 ABSs. After maximum iterations and with more users ($\lambda=10$), the upper limit for accuracy is 80\%, which is 10\% lower than the case of lesser users ($\lambda=5$). Fig.~\ref{g2} shows the cumulative probability curve with upper and lower bounds for a maximum number of users handled by accurately allocated Tier-1 ABSs. With more ABSs assigned to accurate locations, the probability of handling the users also increases. To showcase the performance, the results are evaluated for handling the users after the first iteration. Fig.~\ref{g3} shows that the scenarios with maximum ABSs allocated accurately in the first iteration provide better coverage. Network with $\lambda=5$ provides 83.9\% and 84.9\% better coverage than the networks with $\lambda=8$ and $\lambda=10$, respectively. Finally, the results are recorded for the likelihood of accurately moving and placing ABSs in respective zones. The results in Fig.~\ref{g4} show that the proposed approach can maximize the likelihood of maximum coverage after fewer iterations. Furthermore, the proposed approach is compared with popular algorithms like Hill Myna and Desert Sparrow Optimization (HMADSO)~\cite{Sharma2018} and two versions of Particle Swarm Optimization (PSO)~\cite{eberhart2001swarm} algorithms as shown in Fig.~\ref{fig_g6}. The proposed approach with direct facilitation from the centroid-Voronoi constellations shows 35.7\%, 91.8\%, 88.3\% better convergence in terms of the number of iterations required to accurately map the ABSs in comparison with HMADSO, PSO, and PSO-V, respectively. Here, PSO is operated by using similar centroid-user modelling (Poisson process) as used by the proposed approach, whereas V-PSO is the vector-PSO, which is operated with a velocity-distance variation for global positioning. These results suggest the high convergence of the proposed approach towards an optimal solution. %%Moreover, these results are evidence that it is desirable to consider survivability, coverage, and mobility laws, irrespective of their mechanisms while using ABSs in multiple tiers in association with general TCN. ABS networks enhance the potential of TCN while resolving their performance issues and assisting them with additional services.

\section{Discussions, Lessons Learned and Open Issues}
Over the past few years, many research organizations have identified a tremendous amount of applications for using single tier ABS networks with multiple drones. Some of them have also emphasized the use of HAPS to facilitate the workflow of the network. However, most of them neglected the governing laws of using multi-ABSs in a hierarchy. This article emphasizes the need for operational laws, which include survivability, coverage, and mobility as core components.
\begin{figure}[!ht]
  \centering
  \includegraphics[width=250px]{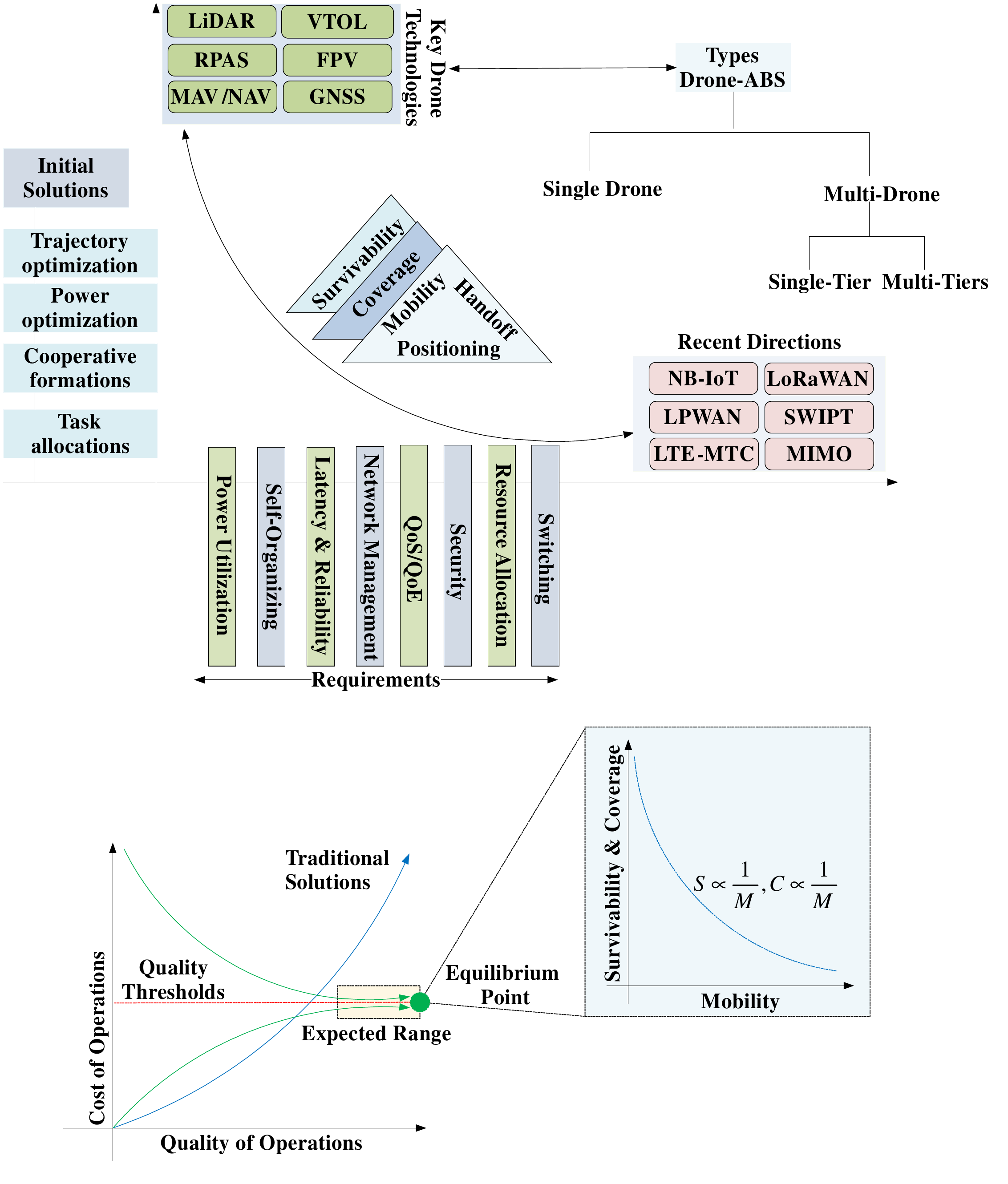}
  \caption{An overview of existing solutions, technologies and research to follow for using ABSs with TCN.}\label{fig_learned}
\end{figure}

The network formulations, result evaluations, and technological discussions provide evidence of enhancement in the functioning of the network by the inclusion of multi-tier ABSs. Such an inclusion allows a better lifetime, better coverage and helps to control the variations due to the high mobility of aerial nodes. \textcolor{black}{Finally, to summarize the understandings of the proposed work and research to follow, an illustration is presented in Fig.~\ref{fig_learned}, which shows the direction of research and issues to be resolved for fully functional utilization of the hierarchical multi-ABSs network. Several challenges to resolve in this direction of research include power utilization, self-organization of drones in tiers, ultra-high reliability and ultra-low latency, mobility management and enhancement of QoS/QoE for the users. Solutions like Narrow-Band IoT (NB-IoT), Long Range Wide Area Network (LoRaWAN), Low Power Wide Area Network (LPWAN), LTE-Machine Type Communications (LTE-MTC), Simultaneous Wireless Information and Power Transfer (SWIPT), Multiple-Input and Multiple-Output (MIMO) or Massive-MIMO, Massive-IoT can be used in the formation of highly survivable ABS networks.} The research issues related to the effective utilization of these technologies and their suitable incorporation for drone communications are still open and have a long way to follow.
\section{Conclusions}
This article focuses on the crucial aspects of survivability, coverage, and mobility laws for the multi-ABSs network. The proposed approach provides an optimal solution for these three factors while maximizing the probability of connectivity and the likelihood of mapping multi-tier ABSs and underlying users. The use of dynamic nodes, such as drones, not only provides the flexibility of operations but also has a considerable impact on the Total Cost of ownership. With hierarchical formations, multi-ABSs can be operated in tiers, which allow significant control over the network and enhance the overall performance. %%Results presented in this article show that the proposed approach accounts for maximizing the accuracy in using multi-tier ABSs according to the geographical area with lesser iterations. In addition, the article also presented details on the several optimization issues, an overview of existing solutions, available technologies, and research to follow for using ABS with Terrestrial Cellular Networks.

\bibliographystyle{ieeetr}
%\bibliography{stin_related}
%\bibliographystyle{IEEEtr}
\bibliography{bibfile.bib}

\footnotesize
\section*{About the Authors}
\noindent\textbf{Vishal Sharma} is working as a Lecturer in the School of Electronics, Electrical Engineering and Computer Science at the Queen's University Belfast (QUB), NI, United Kingdom. His research interests are UAV Communications, 5G and Beyond, Blockchain, Mobile Internet Systems, and CPS-IoT. He attained his PhD degree in Computer Science and Engineering from Thapar University, India. He is a member of the IEEE and ACM. Email: vishal\_sharma2012@hotmail.com\\

\noindent\textbf{Navuday Sharma} is working as an RF Test Development Engineer at Ericsson, Estonia. His research interests are channel modelling, multi-carrier communication, and digital signal processing. He attained his PhD degree in telecommunication engineering from Politecnico di Milano, Italy. He is a member of IEEE. Email: navuday.sharma@ericsson.com\\

\noindent\textbf{Mubashir Husain Rehmani} is working as an Assistant Lecturer in the Department of Computer Science, Munster Technological University (MTU), Cork, Ireland. His research interests are cognitive radio ad hoc networks, smart grid, cognitive radio-based smart grid, wireless energy transfer, flying ad-hoc networks, wireless sensor networks, and mobile ad hoc networks. He received the PhD degree from LIP6, Université of Pierre et Marie Curie (Sorbonne Universités), Paris, France. He is a senior member of IEEE. Email: mshrehmani@gmail.com\\

\noindent\textbf{Haris Pervaiz} is working as a Lecturer in the Department of Computer Science at Lancaster University, UK. His main research interests are Wireless Networking and Communications, including Energy-Efficient Green Communications and Networking, and Cognitive Radio Networks. He received the PhD degree in Communication Systems from Lancaster University, UK. He is a member of IEEE. Email: h.b.pervaiz@lancaster.ac.uk\\
\end{document}